\documentstyle[aps,multicol]{revtex}
\input{epsf}

\begin{document}

\title{Quantum-classical correspondence of  
entropy contours in the transition to chaos}

\author{Raphael Zarum and Sarben Sarkar}
\address{Department of Physics, King's College London, 
Strand, London WC2R 2LS, UK}
\date{\today}

\maketitle

\begin{abstract} 
Von Neumann entropy production rates of the quantised kicked rotor interacting
with an environment are calculated.
A significant correspondence is found between the entropy contours
of the classical and quantised systems.
This is a quantitative tool for describing quantum-classical 
correspondence in the transition to chaos.
\end{abstract}

\bigskip 
\bigskip
\bigskip

\begin{multicols}{2}

\section{Introduction}

The fundamental split between integrable and non-integrable systems in 
classical mechanics has not been comprehensively mirrored in
quantum mechanics \cite{ford91}.
The issue seems to hinge on finding a suitable definition for 
{\it quantum} chaos.
The sensitive dependence to initial conditions that characterises classical 
chaos is wholly understood in terms of trajectories of classical phase space  
points which have no direct quantum analog \cite{schack96a}. 
This has led to two distinct ways of identifying variables to measure
quantum chaos.
One involves investigating various quantum variables that can act as 
signatures of chaos by clearly distinguishing between quantum systems 
whose classical counterparts are integrable and those which 
are non-integrable. 
The implementation of an expanding array of energy spectra properties 
have made this approach highly successful 
(e.g. \cite{reichl92,berry91,haake91,gutzwiller90,tabor89}).
A second approach seeks an intrinsically  quantum definition of quantum chaos 
by investigating the quantum parallels for variables such as 
Lyapunov exponents and various entropy measures which define and quantify 
chaos in classical mechanics 
\cite{ford91,schack96a,entropymethods,schack96b,zurek94,zurek95}.

In this paper we adopt the second approach and develop an original 
technique involving the analysis of entropy production measures to 
reveal a clear correspondence between the quantum and classical formulations 
of a seminal system which has a rich structure of delicately interwoven 
regular and chaotic dynamics - the standard map.
Section II A briefly outlines the main characteristics of the standard map and 
then Section II B describes how classical entropy contours are calculated 
and used to give a comprehensive account of these characteristics.
In a similar fashion, Section III A briefly outlines the quantisation of the 
standard map and then Section III B describes how quantum entropy contours 
are generated through interaction with an environment.
The similarities and differences between the classical and quantum entropy 
contours are explained in Section IV A and Section IV B concludes the paper 
with a discussion on the use of entropy measures to describe quantum chaos.

\section{Classical standard map}
\subsection{Phase space description}

The standard map describes the local behaviour of nonintegrable dynamical 
systems in the separatrix region of non-linear resonances.
The name results from its extensive use in the investigation of chaos, 
especially the mechanisms involved in the transition to global 
chaos in conservative systems. 
It is derived from the kicked rotor model of a one dimensional pendulum
and is a Hamiltonian (area preserving) dynamical system. 
Though the  map has been extensively analysed 
\cite{chirikov79,greene68,greene79,meiss92}, 
here we describe some relevant details.

The standard map can be represented by the equations of motion:

\begin{eqnarray}
p_{n+1} &=& p_n - \frac{K}{2\pi} \sin(2\pi q_n) \nonumber\\
q_{n+1} &=& q_n + p_{n+1}\; (\bmod \; q=1), 
\end{eqnarray}

where $K$ is a real variable which acts as the chaos parameter.
With unit mass and discrete time, $p$ and $q$ have the same dimensions.
The phase space of the map is periodic in $q$, by definition, and as
a special feature of the standard map it is also periodic in $p$, with the
same period 1. 
$K=0$ is the case for a free rotor, a trivial and completely regular map.
In Fig.1(a)-(c), we show the well known sequence of standard map plots for 
increasing values of $K$. 
The map has several axes of symmetry and the unit periodicity in both 
$p$ and $q$ means that only a unit square in phase space need be viewed. 
The interval $ [ -\frac{1}{2},\frac{1}{2} )$ is used for all 
results in this paper.
Fig.1(a) shows the mapping when $K=0.2$. 
Much of the phase space is composed of KAM (Kolmogorov-Arnold-Moser) tori 
\cite{reichl92} stretching horizontally 
from $q=-\frac{1}{2}$ to $q=\frac{1}{2}$ 
and serve to isolate one region of phase space from another.
Periodic orbits in the central resonance can be easily seen, 
as well as a few other dominant non-linear resonances 
(small ellipses) between KAM tori. 

As $K$ is increased, KAM tori that horizontally span the phase 
space are destroyed by resonances and are replaced with smaller KAM islands.
Beyond a critical value at $K \approx 0.97$, the last phase space spanning 
KAM torus is broken and the map becomes globally chaotic (Fig.1(b)). 
Now the remaining resonances are clearly visible as {\it stable islands} 
in a {\it chaotic sea} of trajectories. 
On reaching $K=4$, most structure is wiped out (Fig.1(c)).

\subsection{KS entropy contours}

A positive Lyapunov exponent, which quantifies the exponential divergence 
in time of two closely neighbouring phase space trajectories, 
is a primary definition of classical chaos \cite{oseledec68}. 
For one-dimensional maps such as the standard map the positive Lyapunov 
exponent is equal to the Kolmogorov-Sinai (KS) entropy, $h_{KS}$,
which is a measure of the rate of information production in the system 
\cite{piesin77,benettin76}.
Thus $h_{KS} = 0$ only for completely regular dynamics.
The KS entropy (also called dynamical entropy or metric entropy) of a 
chaotic mapping can be calculated using the formula

\begin{equation}
h_{KS} = \lim_{t\rightarrow\infty} \left( \frac{1}{t} \right) 
\sum_{n=1}^t \log_2 l_n, \label{ks-ent}
\end{equation}

where $l_n = \sqrt{ (\delta p_n)^2 + (\delta q_n)^2 }$ is the changing 
distance between two initially close neighbouring points, 
$(p_0,q_0)$ and $(p_0 + \delta p_0, q_0 + \delta q_0)$, in phase space.
$\delta p$ and $\delta q$ are evolved by iterating a linearised form of the 
chaotic map.
This {\it tangent map} is rescaled after every iteration as follows: 
the $ n^{th}$ iteration of the map produces the values 
$\delta p_n$ and $\delta q_n$ from which $l_n$ is calculated. 
These values are then rescaled to $\; \delta \bar{p}_n = \delta p_n/l_n$ and 
$\; \delta \bar{q}_n = \delta q_n/l_n$ which are fed back into the tangent map 
for the next iteration \cite{chirikov79}.
Use of the base-2 logarithm in (\ref{ks-ent}) allows the entropy to be 
measured in {\it bits} of information.

The standard map is linearised to give its associate tangent map

\begin{equation}
\pmatrix{ \delta p_{n+1}      \cr  \delta q_{n+1} } =
\pmatrix{ 1 & K\cos(2\pi q_n) \cr  1 & 1+K\cos(2\pi q_n) } 
\pmatrix{ \delta p_n          \cr  \delta q_n}, \label{tangent}
\end{equation}

which can be employed in (\ref{ks-ent}) to calculate $h_{KS}$.

The tangent map (\ref{tangent}) clearly shows that the value of $h_{KS}$ 
depends on the initial position in phase space $(p_0,q_0)$ for the standard 
map. 
This is not always the case.
$h_{KS}$ is generally used as a {\it global} measure of the level of chaos in
a given system, but this is only valuable if all chaotic trajectories 
in the system can reach into all regions of its phase space. 
Well known examples of such systems include the cat and baker's 
maps \cite{arnold68}.
However for  $K<1$ the standard map has a predominantly {\it mixed} 
phase space in which different chaotic regions are not connected.
KAM tori act as boundaries so that trajectories originating in one chaotic 
region cannot escape to another.
This isolation inhibits the exponential divergence of chaotic trajectories 
so that the positive Lyapunov exponent, and consequently $h_{KS}$, will vary 
from region to region. 
The kicked top is another example of a well known mixed phase space 
system \cite{haake87}.

To reveal a {\it complete} description of the standard map at a specific $K$ 
in terms of KS entropy, many values of $h_{KS}$ 
corresponding to many initial positions in phase space can be
plotted as a contour map on phase space.
This has been done in Fig.1(d)-(f).
Using (\ref{ks-ent}) and (\ref{tangent}), and setting $t=10^5$ iterations, 
values for $h_{KS}$ were calculated for each point on a 64 x 64 grid 
spanning the same unit of phase space and the same $K$ values as 
in Fig.1(a)-(c). 
Shading intensity reflects the relative sizes of the KS entropy. 
$h_{KS}=0$ is shown as white on the maps while darker and darker shades 
of grey reflect an increasing $h_{KS}$.
The black areas show the largest $h_{KS}$ values corresponding to the 
most chaotic regions of the standard map.
The resemblance between Fig.1(a)-(c) and Fig1.(d)-(f) is striking. 
Stable islands in the classical maps translate to stark white patches in
the contour maps. 
This is because $h_{KS}=0$ for {\it all} periods in non-chaotic dynamics.
The chaotic sea of trajectories in the classical maps are also faithfully 
reproduced as very dark patches of similar shape and size in the
contour maps.
All these correlations indicate that $h_{KS}$ presented in this way can 
comprehensively display all the essential features of the standard map as 
it becomes globally chaotic.

\section{Quantum kicked rotor}
\subsection{Quantization}

The quantised model of the standard map is governed by the Hamiltonian

\begin{equation}
\hat{H}_{stan}(\hat{p},\hat{q},t) =
\frac{\hat{p}^2}{2} -
\frac{K}{4\pi^2} \cos(2\pi\hat{q}) \hspace{-0.1cm}
\sum_{n=-\infty}^{\infty} \hspace{-0.1cm} \delta (t-nT).    \label{hstan}
\end{equation}

This ``kicked rotor'' describes a free particle of unit mass which experiences 
impulses (kicks) at intervals $T$.
Following \cite{casati79} and \cite{berry79}, the kinematics are that of
finite dimensional quantum mechanics with periodic boundary conditions.
Position and momentum space are thus discretized, placing the lattice sites 
at integer values $q_a = p_a = \frac{a}{D}$ 
for $a = -\frac{D}{2},\ldots,\frac{D}{2}-1$. 
The dimension $D$ of Hilbert space is taken as even and, for consistency of 
units, the quantum scale on phase space is taken to be 
$2\pi\hbar = \frac{1}{D}$.
Position and momentum basis kets are denoted by 
$|q_a\rangle$ and $|p_a\rangle$. 

Initial states $|\psi_0\rangle$ are assumed to be  {\it coherent states} 
(minimum-uncertainty states). 
The fiducial initial coherent state 
$|\psi_{0 \{ 00 \} } \rangle = |q_0 , p_0 \rangle$
is defined as the ground state of a special Harper operator \cite{saraceno90},
which can be displaced with the appropriate operators to produce all the other
possible initial coherent states 
$|\psi_{0 \{ ab \} } \rangle = |q_a , p_b \rangle$, i.e.,

\begin{equation}
|\psi_{0 \{ab \} } \rangle = 
\exp \left( \frac{i\pi ab}{D} \right) 
\exp(-2\pi ia\hat{p}) \exp(2\pi ib\hat{q}) 
|\psi_{0 \{ 00 \} } \rangle
\end{equation}

At time $t$, the system can be described by the density operator 
$\rho(t) = |\psi(t)\rangle\langle\psi(t)|$ which changes according to
the evolution equation

\begin{equation}
\rho(t+T) = \hat{U}_s \; \rho(t) \; \hat{U}^{-1}_s,	\label{ev-eq}
\end{equation}

where the kicked rotor unitary evolution operator,

\begin{equation}
\hat{U}_s = \exp \left( \frac{-i\hat{p}^2 T}{2\hbar} \right) 
	    \exp \left( \frac{iKT\cos(2\pi\hat{q})}{4\pi^2\hbar} \right).
\end{equation}

\subsection{Von Neumann entropy contours}

Parallelling the classical case, we look for an entropy measure to reveal 
the dynamics of the quantum system. 
Entropy in quantum statistical mechanics is referred to as von Neumann (vN) 
entropy, $h_{vN}$, (the equivalent measure in classical mechanics is the Gibbs 
entropy) and can be defined in terms of the density matrix $\rho$ of a 
system as

\begin{equation}
h_{vN} = -\mathrm{Tr} \mathnormal{ ( \rho \log_2 \rho ). }
\end{equation}

$h_{vN}$ is a quantative measure of disorder and can be measured in bits.
However, the unitarity of Hamiltonian dynamical evolution dictates that 
$h_{vN}$ remain constant at all times.
The situation can be altered by perturbing the system through the interaction
with an environment.
Averaging over the various possible effects of this environment will then 
lead to an entropy increase $\Delta h_{vN}$ which can then be employed 
to measure the system's chaotic nature. 
This is more than a convienient mathematical construction. 
To produce a quantum kicked rotor in an experimental situation, the free 
particle motion must be periodically opened up to an environment 
to allow the ``kick'' to be introduced \cite{moore95}.
(Though (\ref{ev-eq}) defines the evolution operator for free motion 
experiencing an {\it instantaneous} periodic kick, it is equally valid,
as long as the free motion is not concurrent \cite{tabor89}, for a 
{\it finite} time periodic kick which is what is required to realise this 
experimentally.)
In doing this the environment itself effects the system which naturally
causes the entropy increase required.
 
Thus we choose the environmental coupling to mirror the form of the kick 
in (\ref{hstan}) viz. the $q$ dependence and interaction time.
We also choose the environment model to be a collection of degenerate two-state
atoms with a range of interaction strengths governed by a normal distribution
(this is a generalisation of the class of environments considered by Schack 
and Caves \cite{schack96b}), so that

\begin{equation}
\hat{H}_{int} = 
\frac{\alpha \cos(2\pi\hat{q})}{4\pi^2} \otimes \hspace{-0.1cm}
\sum_{n=-\infty}^{\infty}  \hat{\sigma}_z(n) \delta (t-nT).    \label{hint}
\end{equation}

Thus during the {\it n}th kick the rotor interacts with a single two state 
system with Pauli operator $\hat{\sigma}_z(n)$ and interaction 
strength $\alpha$.
Each of the two-state environment systems is equally likely to be in the
``up'' state 
$| \hspace{-0.1cm} \uparrow \rangle$, 
where 
$\hat{\sigma}_z| \hspace{-0.1cm} \uparrow \rangle = 
| \hspace{-0.1cm} \uparrow \rangle$, 
or in the ``down'' state 
$| \hspace{-0.1cm} \downarrow \rangle$, 
where 
$\hat{\sigma}_z| \hspace{-0.1cm} \downarrow \rangle = 
-| \hspace{-0.1cm} \downarrow \rangle$.
Also, $\alpha$ is drawn from a collection of $M+1$  
independent interaction strengths such that $\alpha = \alpha_j$ for 
$j=-\frac{M}{2},\ldots,0,\ldots,\frac{M}{2}$.  
The distribution $P_{\alpha_j}$ for $\alpha_j$ is the normal distribution
$N(\alpha_0,\alpha_{sd}^2)$. 
 
The combined Hamiltonian for the coupled system and environment is thus

\begin{equation}
\hat{H}_{tot} = \hat{H}_{stan} + \hat{H}_{int}, 
\end{equation}

and the corresponding density operator evolution equation is 

\begin{equation}
\rho(t+T) = \hat{U}_{tot}(\alpha,\lambda) 
\; \rho(t) \; 
\hat{U}^{-1}_{tot}(\alpha,\lambda),
\end{equation}

where the combined evolution operator,

\begin{equation}
\hat{U}_{tot}(\alpha,\lambda) 
= \exp \left( \frac{-i \alpha \lambda T\cos(2\pi\hat{q})}
{4\pi^2\hbar} \right) \hat{U}_s,
\end{equation}

with $\lambda \in \{-1,1\}$ is the result of measuring the two state 
environment after each interval to determine whether it is an up or 
down state. 
(As before, this same operator would result if (\ref{hint}) was turned on 
for the {\it finite} time required for an experimental realisation of this 
system.) 
The effect of this environmental coupling is to produce a multiple 
stochastic perturbation at the end of each time interval.
After each interval, there are $2M+2$ different measurement 
results leading to $2M+2$ possible pure states for the system.
Averaging over all these possible outcomes in the position basis produces
the density operator evolution equation

\begin{eqnarray}
 &\overline{\rho}_{xy}(t+T)& \; \equiv \;  
\langle x| \; \overline{\rho}(t+T) \; | y \rangle  \nonumber\\
 &=& \hspace{-0.3cm}
\sum_{j=1}^{M+1} \frac{P_{\alpha_j}}{2} 
\sum_{ \lambda = -1,1} 
\hspace{-0.2cm} 
\langle x | \; \hat{U}_{tot}(\alpha_j,\lambda) 
\; \overline{\rho}(t) \;  
\hat{U}^{-1}_{tot}(\alpha_j,\lambda) | y \rangle  \nonumber\\
 &=& F(x,y) \; \langle x| \; \hat{U}_s 
\; \overline{\rho}(t) \; 
\hat{U}^{-1}_s \; |y \rangle,
\end{eqnarray}

where

\begin{equation}
F(x,y) = \sum_{j=1}^{M+1} 
P_{\alpha_j} \cos  
\left( 2 \alpha_j D \pi  
\sin \frac{\pi (x+y)}{D}
\sin \frac{\pi (x-y)}{D} 
\right)
\end{equation}

now contains all the perturbation effects due to the environment.
This causes a vN entropy increase which can be determined by 
tracing over the system so that,

\begin{equation}
\Delta h_{vN} (nT) = -\mathrm{Tr} 
( \mathnormal{\overline{\rho}(nT) \log_2 \overline{\rho}(nT)}),
\end{equation}

where $\overline{\rho} (nT)$ is the average density matrix of the 
system after $n$ time intervals.


One final step will allow us to see the quantum chaotic dynamics.
Zurek and Paz \cite{zurek94} have conjectured that for an {\it open} quantum
system with minimal dissipation which displays classical chaos, 
the {\it rate} of vN entropy production, $\tilde{h}_{vN}$,
of its quantum analogue, after an initial decoherence time, $t_d$,
will rise to a maximum value which is solely dependent on the sum of 
its positive Lyapunov exponents. 
This will continue to be the case until the system begins to approach 
equilibrium when $\tilde{h}_{vN}$ will slowly decrease reaching zero at 
time $t_{eqm}$. 
In contrast, the entropy production rate of the quantum analogue of a 
regular systems will asymptotically tend to zero well before $t_{eqm}$.
Applying this to the standard map, $\tilde{h}_{vN}$ for the quantised 
system interacting with an environment should be comparable to the 
KS entropy of its classical (unperturbed) counterpart.
Thus for $t_d < nT \ll t_{eqm}$,

\begin{equation}
\tilde{h}_{vN} \approx
\frac{\Delta h_{vN}(nT) - \Delta h_{vN}((n-1)T)}{T} \approx h_{KS}.
\end{equation} 
 
A uniformly spaced 64 x 64 grid of initial 
coherent states (corresponding to an even spread over unit phase space)
were numerically evolved in time. 
The maximum value of $\tilde{h}_{vN}$ for each evolution was plotted on 
a contour map in a similar fashion to the classical case.
Fig.1(g)-(i) displays the results for the same three values of $K$ 
with $D=256$,
$\alpha_0 = 0.001$,
$\alpha_{sd} = 0.2\alpha_0$, $M=100$ and $T=1$.

\section{Discussion}
\subsection{Quantum-classical correspondence}

There are remarkable similarities between Fig.1(d)-(f) and Fig.1(g)-(i). 
For the same $K$ values, the size and location of the various stable islands is
analogous, dark patches are prevelant in the heavily chaotic regions, 
the axes of symmetry are consistent and the overall complexity of the 
dynamics is clearly visible in both.

There are also differences.
The quantum contour maps are generally much smoother than their classical 
counterparts.
This is because each initial coherent state in the quantum system has a 
support area causing their evolution to imitate that of a {\it density} 
of points on phase space.
Thus neighbouring coherent states will fail to achieve dramatically different 
rates of vN entropy production.
Increasing $D$ reduces the supports of the initial coherent states
as well as reducing the overlap between neighbouring states.
It was found that this led to a reduction in the smoothness of the 
quantum contour maps making them more greatly resemble the classical 
contour maps.

The process of calculating these entropy contour maps was repeated with
variations to $N(\alpha_0,\alpha_{var}^2)$. 
For $\alpha_0,\alpha_{sd} \ll K$, when any entropy increase is due
primarily to the chaotic dynamics of the system and not the interaction, 
similar results were achieved.

\subsection{Conclusion}

We have demonstrated that an entropy based approach allows
classical chaotic dynamics to be accurately measured in a trajectory 
independent way which in turn makes it eminently suitable to 
measure \cite{sarkar88} and analyse the corresponding 
quantum chaotic dynamics. 
We have also given numerical support to the Zurek and Paz conjecture 
in a chaotic system which folds phase space, a characteristic that their 
conjecture did not directly take into account.

Entropy measures for diagnosing chaotic dynamics can also be employed in 
other maps.
The sawtooth map \cite{percival87} (which becomes Arnold's cat map 
\cite{arnold68} for a specific value of the chaos parameter $K$) does not have
a mixed phase space so entropy contours would be of little interest.
However, correspondence {\it can} be investigated by comparing quantum and
classical entropy measures for a {\it range} of $K$ values.
Even more interesting is the kicked top \cite{haake87} which is described 
by a map on the unit sphere.
Like the standard map, it has a predominantly mixed phase space for lower 
values of its chaos parameter making it ideal for comparing classical and
quantum entropy contours.
The kicked top also has a special ``order-within-chaos'' 
\cite{constantoudis97} feature.
In general $h_{KS}$ increases monotonically with the chaos parameter $K$ of a 
given map. 
However, the kicked top has islands of stability reappearing for specific 
higher $K$ values when the map is already globally chaotic.
This leads to an intricate relationship between $h_{KS}$ and $K$ which can be
compared to numerical results for the corresponding $\tilde{h}_{vN}$.     
We will discuss the results for these maps elsewhere.

Finally, KS entropy is information-theoretically defined as the rate of 
production of Shannon entropy (also called the Shannon information 
measure) \cite{shannon48}.
Thus, within well defined parameters, we have shown that the quantum-classical 
correspondence of chaotic dynamical systems may be realised by viewing the 
Shannon entropy production rate as the classical measure 
corresponding to the quantum measure of the von Neumann entropy 
production rate.
These results provide a new diagnostic for investigating the chaotic nature 
of quantum systems.

\section*{Acknowledgements}
We are grateful to the ESPRC for their financial support.
RZ would like to thank  R\"{u}diger Schack for useful discussions.


\end{multicols}
\onecolumn


\begin{figure*}[htb]
\vspace*{-0.5cm}
\hspace*{-1.1cm}
\epsfxsize=7.25cm
\epsfysize=7.25cm
\epsffile{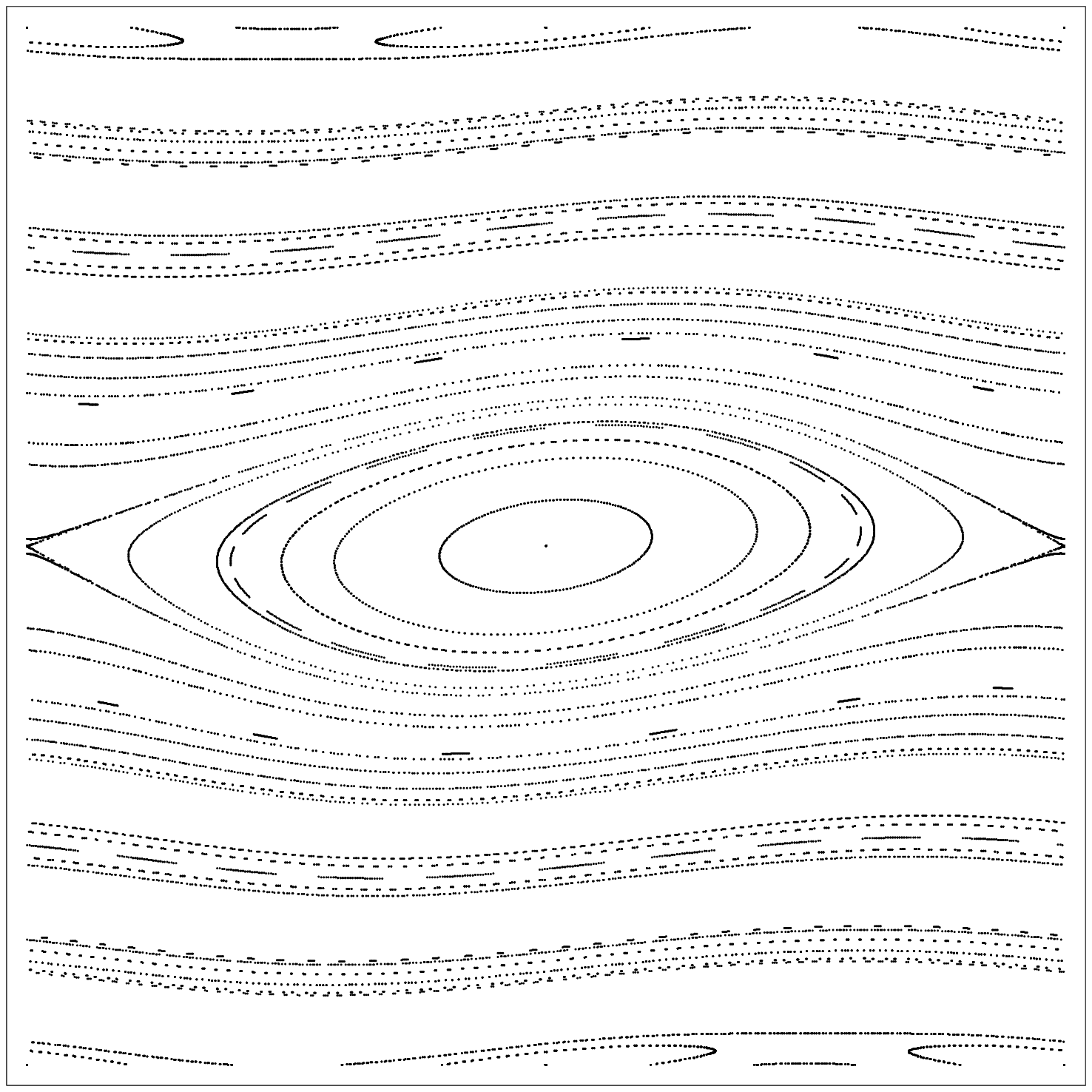}

\vspace*{-7.25cm}
\hspace*{4.9cm}
\epsfxsize=7.25cm
\epsfysize=7.25cm
\epsffile{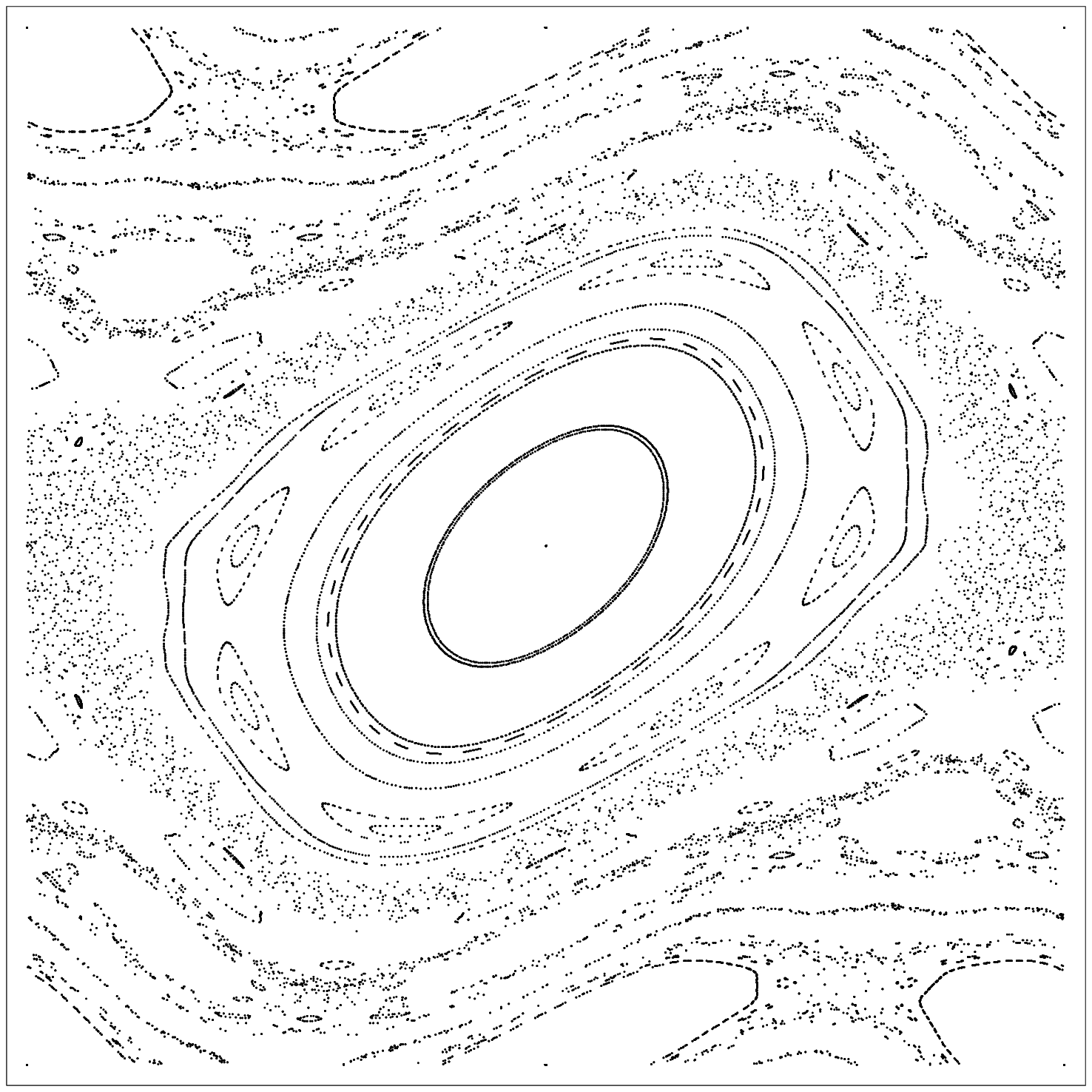}

\vspace*{-7.25cm}
\hspace*{10.9cm}
\epsfxsize=7.25cm
\epsfysize=7.25cm
\epsffile{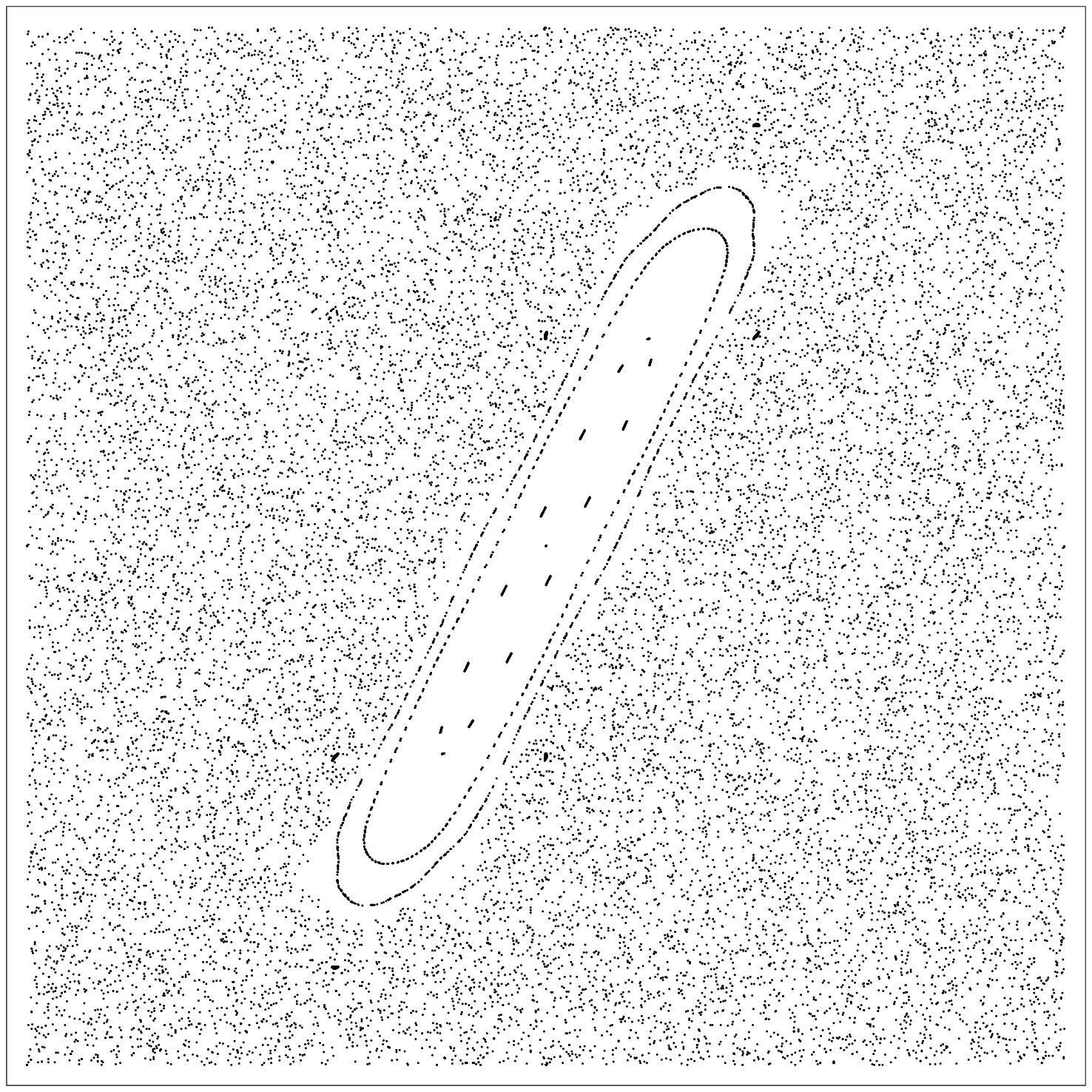}

\vspace{-1.1cm}
\hspace{2.2cm}(a) K=0.2
\hspace{4.3cm}(b) K=1.0
\hspace{4.3cm}(c) K=4.0

\vspace*{0.5cm}
\hspace*{0cm}
\epsfxsize=5.5cm
\epsfysize=5.5cm
\epsffile{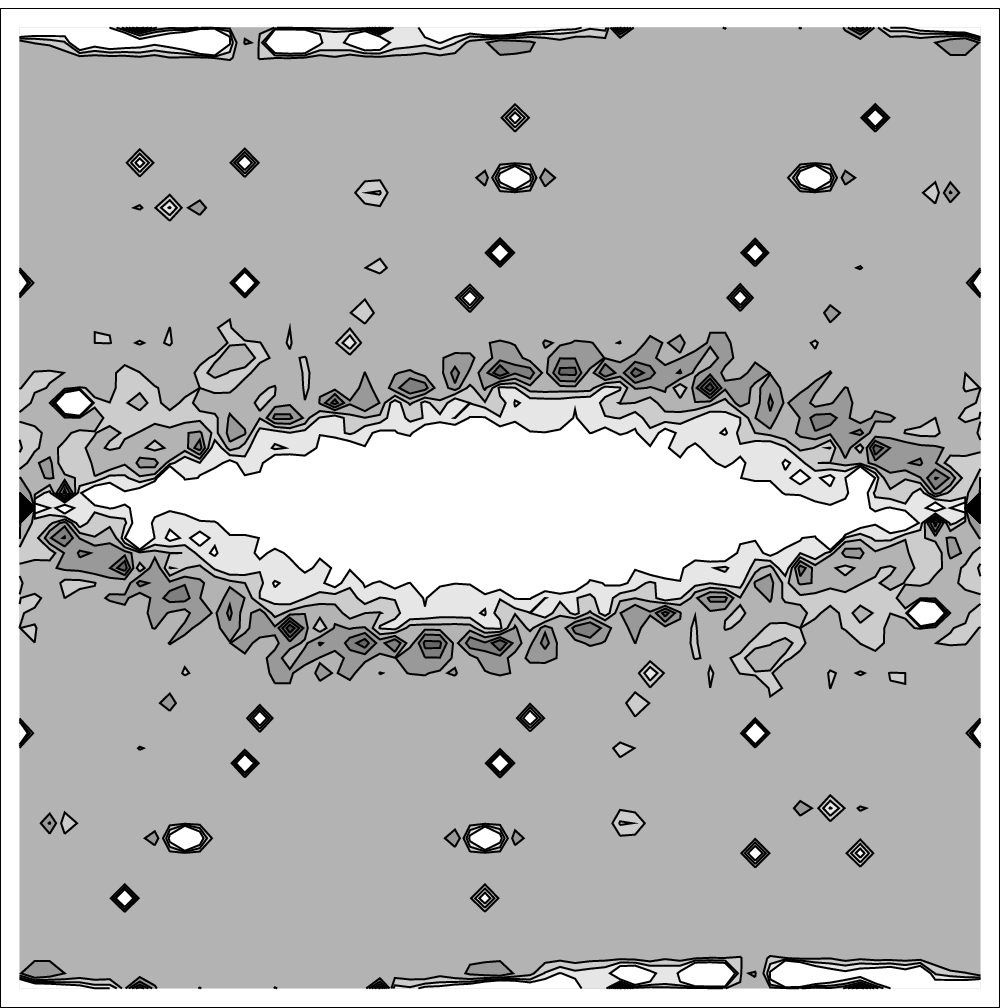}

\vspace*{-5.5cm}
\hspace*{6cm}
\epsfxsize=5.5cm
\epsfysize=5.5cm
\epsffile{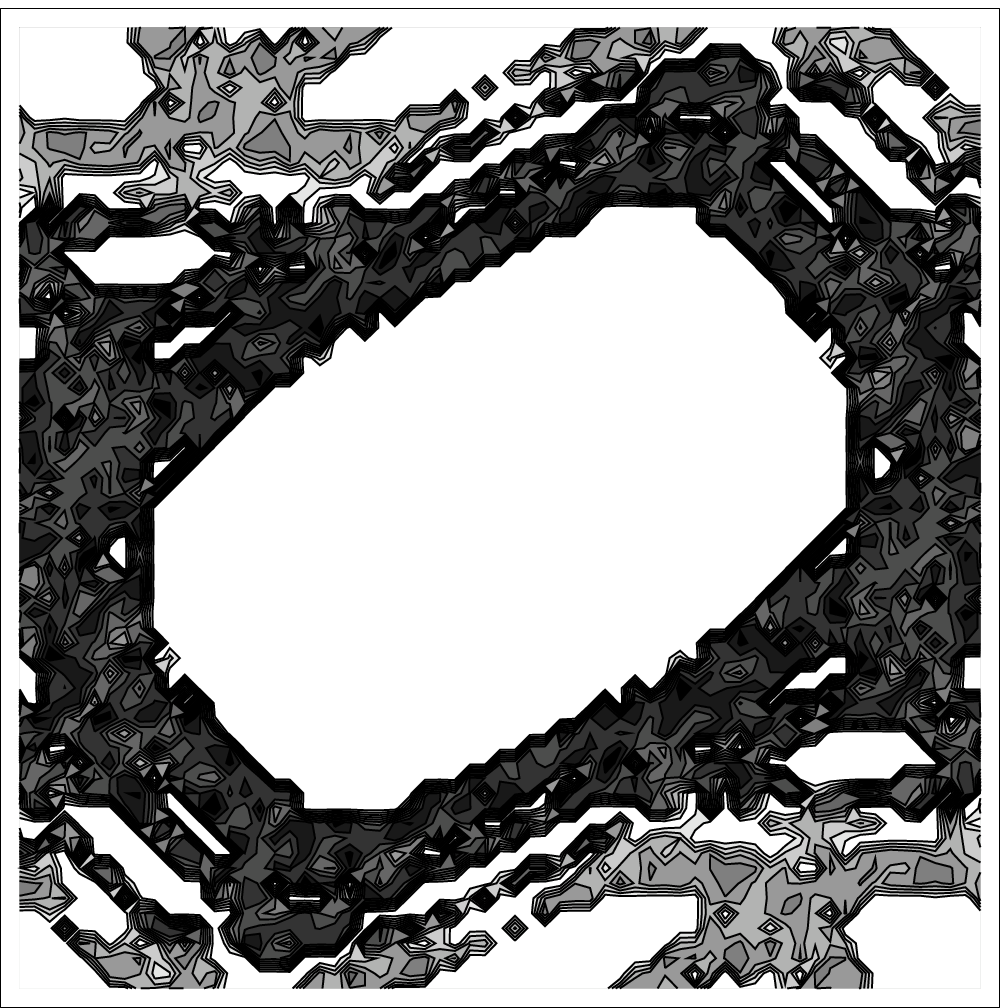}

\vspace*{-5.5cm}
\hspace*{12cm}
\epsfxsize=5.5cm
\epsfysize=5.5cm
\epsffile{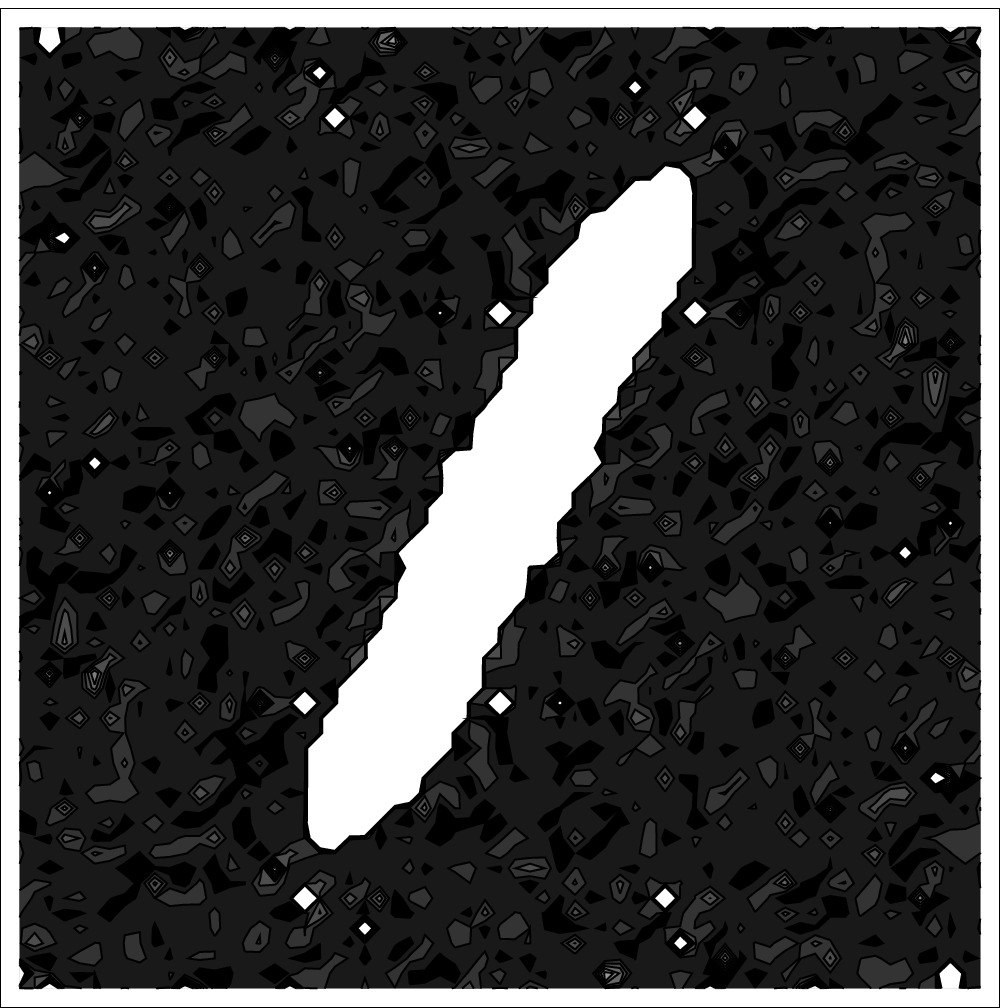}

\vspace{0.1cm}
\hspace{2.2cm}(d) K=0.2
\hspace{4.3cm}(e) K=1.0
\hspace{4.3cm}(f) K=4.0

\vspace*{0.5cm}
\hspace*{0cm}
\epsfxsize=5.5cm
\epsfysize=5.5cm
\epsffile{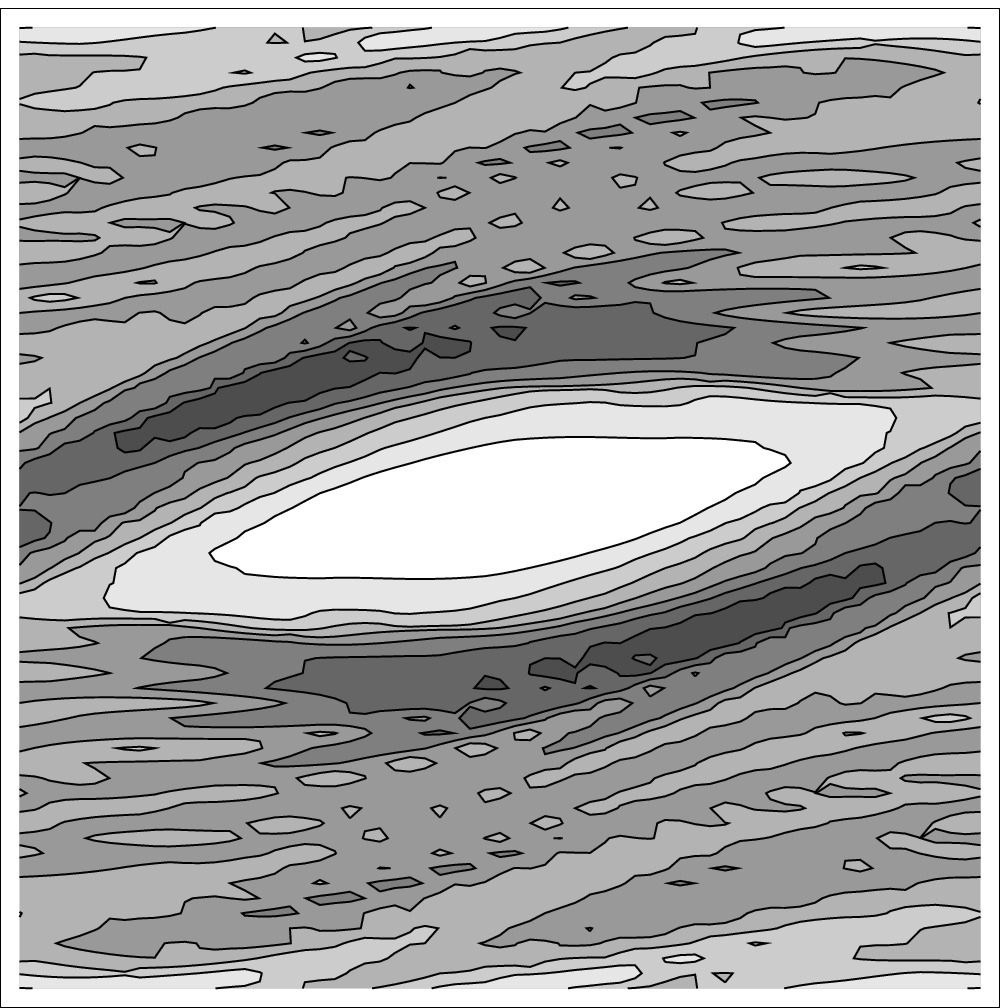}

\vspace*{-5.5cm}
\hspace*{6cm}
\epsfxsize=5.5cm
\epsfysize=5.5cm
\epsffile{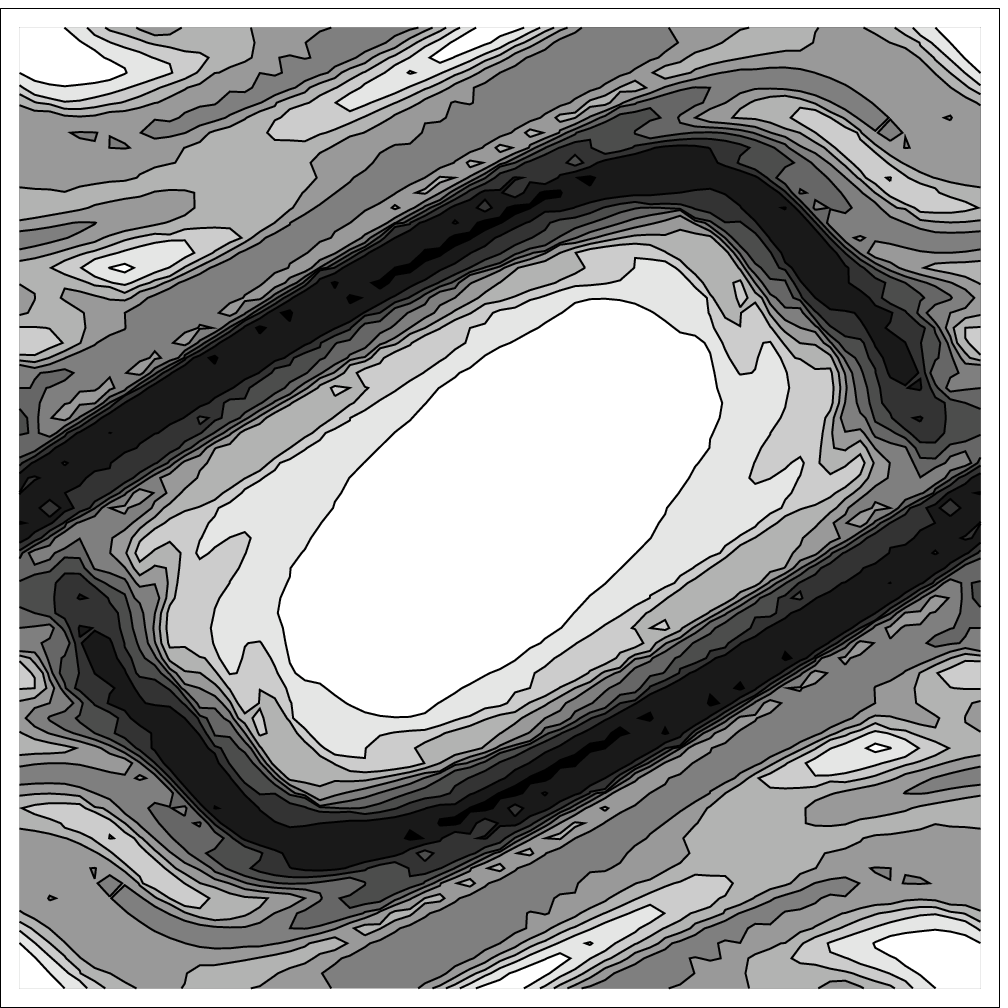}

\vspace*{-5.5cm}
\hspace*{12cm}
\epsfxsize=5.5cm
\epsfysize=5.5cm
\epsffile{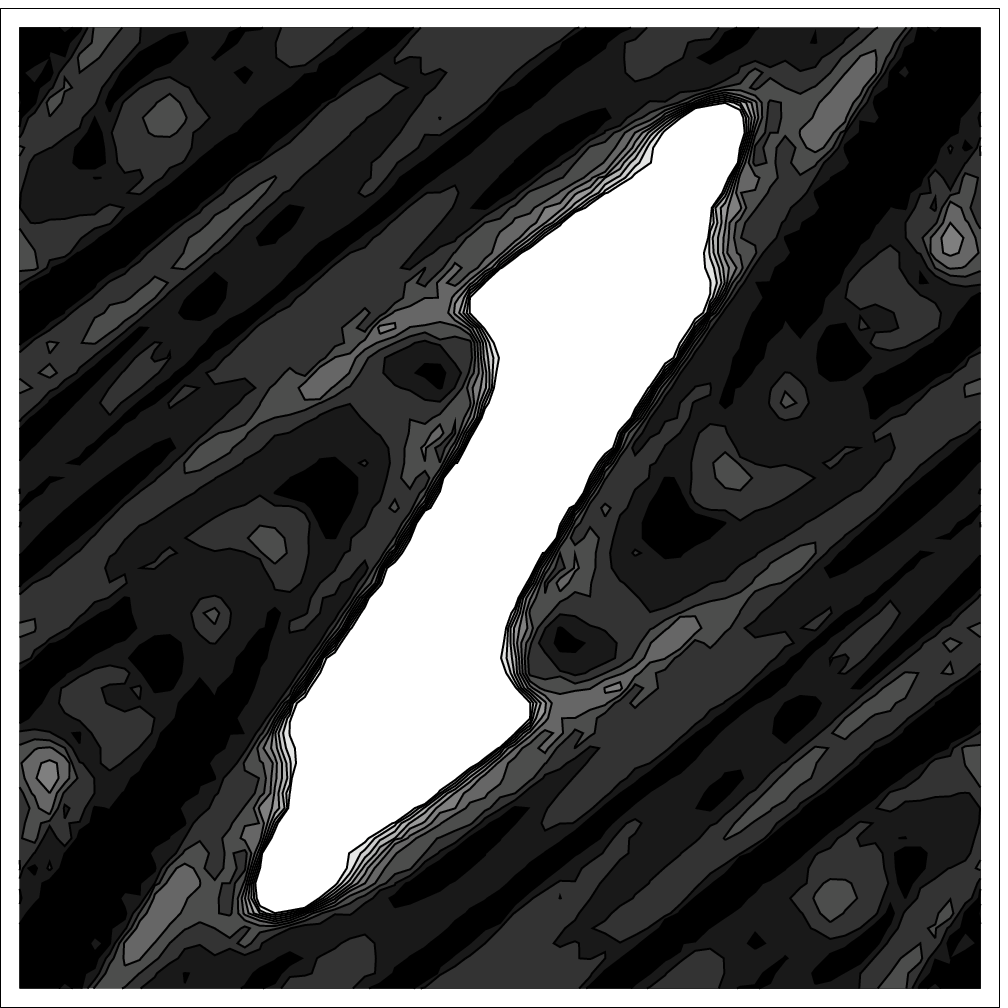}

\vspace{0.1cm}
\hspace{2.2cm}(g) K=0.2
\hspace{4.3cm}(h) K=1.0
\hspace{4.3cm}(i) K=4.0

\vspace{0.4cm}

\caption{Quantum-classical correspondence in the standard map. 
Three values of the chaos parameter $K$ are shown for each set of maps 
clearly showing that correspondence is accurately maintained during the 
transition to global chaos. 
(a)-(c) Classical map in unit phase space bounded on the interval 
$[-\frac{1}{2},\frac{1}{2})$ for both $q$ and $p$.
(d)-(f) Contour plot of KS entropy in unit phase space for the classical map.
(g)-(i) Contour plot of von Neumann entropy production rate in unit phase 
space for the quantum kicked rotor interacting with an environment.}

\end{figure*}


\begin{thebibliography}{99}

\bibitem {ford91} J. Ford, G. Mantica and G.H. Ristow,
Physica D {\bf50}, 493 (1991).

\bibitem {schack96a} R. Schack and C.M. Caves,
Phys. Rev. E {\bf53}, 3387 (1996).

\bibitem {reichl92} L.E. Reichl,
{\it The Transition to Chaos in Conservative Classical Systems: 
Quantum Manifestations}
(Springer-Verlag, Berlin, 1992).

\bibitem {berry91} M.V. Berry, in
{\it Chaotic Behaviour in Deterministic Systems, Les Houches XXXVI},
edited by G. Iooss, R.H.G. Hellman and R. Stora
(North-Holland Amsterdam, 1991).

\bibitem{haake91} 
F. Haake, {\it Quantum Signatures of Chaos},
(Springer-Verlag, Berlin, 1991).

\bibitem{gutzwiller90}
M.C. Gutzwiller, {\it Chaos in Classical and Quantum Mechanics}
(Springer-Verlag, New York, 1990).

\bibitem{tabor89} M. Tabor, 
{\it Chaos and Integrability in Nonlinear Dynamics: An Introduction},
(Wiley, New York, 1989).

\bibitem{entropymethods} 
W. S{\l}omczy\'{n}ski and K. \.{Z}yczkowski,
J. Math. Phys. {\bf35}, 5674 (1994);
S. Klimek and A. Le\'{s}niewski,
Ann. Phys. (N.Y.) {\bf248}, 173 (1996).

\bibitem {schack96b} R. Schack and C.M. Caves,
Phys. Rev. E {\bf53}, 3257 (1996).

\bibitem {zurek94} W.H. Zurek and J.P. Paz,
Phys. Rev. Lett. {\bf72}, 2508 (1994).

\bibitem {zurek95} W.H. Zurek and J.P. Paz,
Physica D {\bf83}, 300 (1995).

\bibitem {chirikov79} B.V. Chirikov, 
Phys. Rep. {\bf52}, 263 (1979).

\bibitem {greene68} J.M. Greene, 
J. Math. Phys. {\bf9}, 760 (1968).

\bibitem {greene79} J.M. Greene, 
J. Math. Phys. {\bf20}, 1183 (1979).

\bibitem {meiss92} J.D. Meiss, 
Rev. Mod. Phys. {\bf63} 795 (1992).

\bibitem {oseledec68} V.I. Oseledec,
Trans. Mosc. Math. Soc. {\bf19}, 197 (1968);
J.P. Eckmann and D. Ruelle,
Rev. Mod. Phys. {\bf57}, 617 (1985).

\bibitem {piesin77} Ya. B. Pesin,
Russ. Math. Surveys {\bf32}, 55 (1977).

\bibitem {benettin76} G. Benettin, L. Galgani and J.M. Strelcyn,
Phys. Rev. A{\bf14}, 2338 (1976).

\bibitem {arnold68} V.I. Arnold and A. Avez,
{\it Ergodic Problems of Classical Mechanics}
(Benjamin, New York, 1968)

\bibitem {haake87} F. Haake, M. Ku\a's and R. Scharf,
Z. Phys. B {\bf 65}, 381 (1987). 

\bibitem {casati79} G. Casati, B.V. Chirikov, F.M. Izraelev and J. Ford, in
{\it Lecture Notes in Physics; 93}, edited by G. Casati and J. Ford 
(Springer-Verlag, 1979).

\bibitem {berry79} M.V. Berry, N.L. Balaszs, M. Tabor and A. Voros,
Ann. Phys. (N.Y.) {\bf122}, 26 (1979).

\bibitem {saraceno90} M. Saraceno,
Ann. Phys. (N.Y.) {\bf199}, 37 (1990).

\bibitem {moore95} F.L. Moore, J.C. Robinson, C.F. Bharucha, B. Sundaram
and M.G. Raizen,
Phys. Rev. Lett. {\bf75}, 4598 (1995).

\bibitem {sarkar88} S. Sarkar and J.S. Satchell,
Physica D {\bf 29}, 343 (1988).

\bibitem {percival87} I.C. Percival and F. Vivaldi,
Physica D {\bf 27}, 373 (1987).

\bibitem {constantoudis97} V. Constantoudis and N. Theodorakopoulos,
Phys. Rev. E {\bf 56}, 5189 (1997).

\bibitem {shannon48} C.E. Shannon and W. Weaver,
{\it The Mathematical Theory of Communication} 
(University of Illinois Press, 1949).

\end{thebibliography}
\end{document}